
\input phyzzx   
%
%
%
\newcount        \ObjClass
\chardef\ClassNum       = 0
\chardef\ClassMisc      = 1
\chardef\ClassEqn       = 2
\chardef\ClassRef       = 3
\chardef\ClassFig       = 4
\chardef\ClassTbl       = 5
\chardef\ClassThm       = 6
\chardef\ClassStyle     = 7
\chardef\ClassDef       = 8
\edef\NumObj    {\ObjClass = \ClassNum   \relax}
\edef\MiscObj   {\ObjClass = \ClassMisc  \relax}
\edef\EqnObj    {\ObjClass = \ClassEqn   \relax}
\edef\RefObj    {\ObjClass = \ClassRef   \relax}
\edef\FigObj    {\ObjClass = \ClassFig   \relax}
\edef\TblObj    {\ObjClass = \ClassTbl   \relax}

\edef\StyleObj  {\ObjClass = \ClassStyle \relax}
\edef\DefObj    {\ObjClass = \ClassDef   \relax}
%
%
\def\gobble      #1{}%
\def\trimspace   #1 \end{#1}%
\def\ifundefined #1{\expandafter \ifx \csname#1\endcsname \relax}%
\def\trimprefix  #1_#2\end{\expandafter \string \csname #2\endcsname}%
\def\skipspace #1#2#3\end%
    {%
    \def \temp {#2}%
    \ifx \temp\space \skipspace #1#3\end
    \else \gdef #1{#2#3}\fi
    }%
\def\stylename#1{\expandafter\expandafter\expandafter
    \gobble\expandafter\string\the#1}
\ifundefined {protect} \let\protect=\relax \fi
\def\checkchapterlabel%
    {%
    \catcode`\@=11
    {\protect\expandafter\if\chapterlabel\noexpand\relax
        \global\let\chapterlabel=\relax \fi}
    \catcode`\@=12
    }%
\begingroup
\catcode`\<=1 \catcode`\{=12
\catcode`\>=2 \catcode`\}=12
\xdef\LBrace<{>%
\xdef\RBrace<}>%
\endgroup
%
%
\newcount\equanumber \equanumber=0
\newcount\eqnumber   \eqnumber=0
\def\(#1)%
        {%
        \ifnum \equanumber<0 \eqnumber=-\equanumber
            \advance\eqnumber by -1 \else
            \eqnumber=\equanumber\fi
        \ifmmode\ifinner(\eqnum {#1})\ifdraft{\rm[#1]}\fi%
        \else \expandafter\ifx\csname Eq_#1\endcsname \relax
        \eqno(\eqnum {#1})\ifdraft{\rm[#1]}\fi \else(\eqnum {#1})\fi\fi
        \else(\eqnum {#1})\fi\ifnum%
            \equanumber<0 \global\equanumber=-\eqnumber\global\advance%
            \equanumber by -1\else\global\equanumber=\eqnumber\fi}%
\def\eqnum #1%
    {%
    \LookUp Eq_#1 \using\eqnumber\neweqnum
    {\rm\label}%
    }%
\def\neweqnum #1#2%
    {%
    \checkchapterlabel%
    {\protect%
        \xdef\eqnoprefix{\ifundefined{chapterlabel}\else\chapterlabel.\fi}}%
    \ifmmode \xdef #1{\eqnoprefix #1}%
        \else\message{Undefined equation \string#1 in non-math mode.}%
             \string#1 \let #1=\relax%
             \global\advance \eqnumber by -1 %
        \fi%
    \EqnObj \SaveObject{#1}{#2}%
    }%
\everydisplay = {\expandafter \let\csname Eq_\endcsname=\relax
                 \expandafter \let\csname Eq_?\endcsname=\relax}%
%
%
\newcount\tablecount \tablecount=0
\def\Table  #1{Table~\tblnum {#1}}%
\def\tblnum #1{\TblObj \LookUp Tbl_#1 \using\tablecount
        \SaveObject \label\ifdraft [#1]\fi}%
\def\tbldef #1{\TblObj \SaveContents {Tbl_#1}}%
\def\inserttable #1#2#3%
    {%
    \tbldef {#1}{#3}\goodbreak%
    \midinsert
        \smallskip
        \hbox{\singlespace
                   \hskip 0.4in
                   \vtop{\parshape=2 0pt 361pt 58pt 303pt
                        \noindent{\bf\Table{#1}}.\enspace #3}
                   \hfil}
        #2
        \smallskip%
    \endinsert
    }%
%
%
\newcount\figurecount \figurecount=0
\def\Figure #1{Figure~\fignum {#1}}%
\def\fignum #1{\FigObj \LookUp Fig_#1 \using\figurecount
     \SaveObject \label\ifdraft [#1]\fi}%
\def\figdef #1{\FigObj \SaveContents {Fig_#1}}%
\def\figlist  {\FigObj \ListObjects}%
\def\insertfigure #1#2#3%
    {%
    \figdef {#1}{#3}%
    \midinsert
        \smallskip
        #2
        \hbox{  \singlespace
                \hskip 0.4in
                \vtop{\parshape=2 0pt 361pt 65pt 296pt
                      \noindent{\bf\Figure{#1}}.\enspace #3}
                \hfil}
        \smallskip%
    \endinsert
    }%
%
%
\newcount\theoremcount \theoremcount=0
\def\fbf#1{\expandafter\ifx\csname Thm_#1\endcsname\relax \bf\fi}
%
%
%
%
%
%
%
%
%
%
%
%
%
\newcount\referencecount \referencecount=0
\newcount\refsequence   \refsequence=0
\newcount\lastrefno     \lastrefno=-1
\def\refsymbol#1{[\refrange#1-\end]}%
\def\[#1]%
        {\refsymbol{#1}}
\def\^[#1]#2{
        \if.#2\rlap.\attach{\refsymbol{#1}}\let\refendtok=\relax%
        \else\if,#2\rlap,\attach{\refsymbol{#1}}\let\refendtok=\relax%
        \else\attach{\refsymbol{#1}}\let\refendtok=#2\fi\fi%
        \discretionary{}{}{}\refendtok}%
\def\refrange #1-#2\end%
    {%
    \refnums #1,\end
    \def \temp {#2}%
    \ifx \temp\empty \else -\expandafter\refrange \temp\end \fi
    }%
\def\refnums #1,#2\end%
    {%
    \def \temp {#1}%
    \ifx \temp\empty \else \skipspace \temp#1\end\fi
    \ifx \temp\empty
        \ifcase \refsequence
            \or\or ,\number\lastrefno
            \else  -\number\lastrefno
        \fi
        \global\lastrefno = -1
        \global\refsequence = 0
    \else
        \RefObj \edef\temp {Ref_\temp\space}%
        \expandafter \LookUp \temp \using\referencecount\SaveObject
        \global\advance \lastrefno by 1
        \edef \temp {\number\lastrefno}%
        \ifx \label\temp
            \global\advance\refsequence by 1
        \else
            \global\advance\lastrefno by -1
            \ifcase \refsequence
                \or ,%
                \or ,\number\lastrefno,%
            \else   -\number\lastrefno,%
            \fi
            \label
            \global\refsequence = 1
            \ifx\suffix\empty
                \global\lastrefno = \label
            \else
                \global\lastrefno = -1
            \fi
        \fi
        \refnums #2,\end
    \fi
    }%
%
%
%
%
\def\refdef #1{\RefObj \SaveContents {Ref_#1}}%
\def\reflist  {\RefObj \ListObjects}%
%
%
%
%
\newif\ifSaveFile
\newif\ifnotskip
\newwrite\SaveFile
\let\IfSelect=\iftrue
\edef\savefilename {\jobname.aux}%
\def\Def#1#2%
    {%
    \expandafter\gdef\noexpand#1{#2}%
    \DefObj \SaveObject {#2}{\expandafter\gobble\string#1}%
}%
\def\savestate%
    {%
    \ifundefined {chapternumber} \else
        \NumObj \SaveObject {\number\chapternumber}{chapternumber} \fi
        \ifundefined {appendixnumber} \else
        \NumObj \SaveObject {\number\appendixnumber}{appendixnumber} \fi
    \ifundefined {sectionnumber} \else
        \NumObj \SaveObject {\number\sectionnumber}{sectionnumber} \fi
    \ifundefined {pagenumber} \else
        \advance\pagenumber by 1
        \NumObj \SaveObject {\number\pagenumber}{pagenumber}%
        \advance\pagenumber by -1 \fi
    \NumObj \SaveObject {\number\equanumber}{equanumber}%
    \NumObj \SaveObject {\number\tablecount}{tablecount}%
    \NumObj \SaveObject {\number\figurecount}{figurecount}%
    \NumObj \SaveObject {\number\theoremcount}{theoremcount}%
    \NumObj \SaveObject {\number\referencecount}{referencecount}%
    \checkchapterlabel
    \ifundefined {chapterlabel} \else
        {\protect\xdef\chaplabel{\chapterlabel}}
        \MiscObj \SaveObject \chaplabel {chapterlabel} \fi
    \ifundefined {chapterstyle} \else
        \StyleObj \SaveObject {\stylename{\chapterstyle}}{chapterstyle} \fi
    \ifundefined {appendixstyle} \else
        \StyleObj \SaveObject {\stylename{\appendixstyle}}{appendixstyle}\fi
}%
\def\Contents #1{\ObjClass=-#1 \SaveContents}%
\def\Define #1#2#3%
    {%
    \ifnum #1=\ClassNum
        \global \csname#2\endcsname = #3 %
    \else \ifnum #1=\ClassStyle
        \global \csname#2\endcsname\expandafter=
        \expandafter{\csname#3\endcsname} %
    \else \ifnum #1=\ClassDef
        \expandafter\gdef\csname#2\endcsname{#3} %
    \else
        \expandafter\xdef \csname#2\endcsname {#3} \fi\fi\fi %
    \ObjClass=#1 \SaveObject {#3}{#2}%
    }%
\def\SaveObject #1#2%
    {%
    \ifSaveFile \else \OpenSaveFile \fi
    \immediate\write\SaveFile
        {%
        \noexpand\IfSelect\noexpand\Define
        {\the\ObjClass}{#2}{#1}\noexpand\fi
        }%
    }%
\def\SaveContents #1%
    {%
    \ifSaveFile \else \OpenSaveFile \fi
    \BreakLine
    \SaveLine {#1}%
    }%
\begingroup
    \catcode`\^^M=\active %
\gdef\BreakLine %
    {%
    \begingroup %
    \catcode`\^^M=\active %
    \newlinechar=`\^^M %
    }%
\gdef\SaveLine #1#2%
    {%
    \toks255={#2}%
    \immediate\write\SaveFile %
        {%
        \noexpand\IfSelect\noexpand\Contents
        {-\the\ObjClass}{#1}\LBrace\the\toks255\RBrace\noexpand\fi%
        }%
    \endgroup %
    }%
\endgroup
\def\ListObjects #1%
    {%
    \ifSaveFile \CloseSaveFile \fi
    \let \IfSelect=\GetContents \ReadFileList #1,\savefilename\end
    \let \IfSelect=\IfDoObject  \input \savefilename
    \let \IfSelect=\iftrue
    }%
\def\ReadFileList #1,#2\end%
    {%
    \def \temp {#1}%
    \ifx \temp\empty \else \skipspace \temp#1\end \fi
    \ifx \temp\empty \else \input #1 \fi
    \def \temp {#2}%
    \ifx \temp\empty \else \ReadFileList #2,\end \fi
    }%
\def\GetContents #1#2#3%
    {%
    \notskipfalse
    \ifnum \ObjClass=-#2
        \expandafter\ifx \csname #3\endcsname \relax \else \notskiptrue \fi
    \fi
    \ifnotskip \expandafter \DefContents \csname #3_\endcsname
    }%
\def\DefContents #1#2{\toks255={#2} \xdef #1{\the\toks255}}%
\def\IfDoObject #1#2%
    {%
    \notskipfalse \ifnum \ObjClass=#2 \notskiptrue\fi \ifnotskip \DoObject
    }%
\def\DoObject #1#2%
    {%
    \ifnum \ObjClass = \ClassTbl        \par\noindent Table~#2.
    \else \ifnum \ObjClass = \ClassFig  \par\noindent Figure~#2.
    \else \item {#2.}
    \fi\fi
    \ifdraft\edef\temp {\trimprefix #1\end}[\expandafter\gobble \temp]~\fi
    \expandafter\ifx \csname #1_\endcsname \relax
        \ifdraft\relax\else\edef\temp {\trimprefix #1\end}%
        [\expandafter\gobble \temp]\fi%
    \else
        \csname #1_\endcsname
    \fi
    }%
\def\OpenSaveFile   {\immediate\openout\SaveFile=\savefilename
                     \global\SaveFiletrue}%
\def\CloseSaveFile  {\immediate\closeout\SaveFile \global\SaveFilefalse}%
\OpenSaveFile\CloseSaveFile 
%
%
\def\LookUp #1 #2\using#3#4%
    {%
    \expandafter \ifx\csname#1\endcsname \relax
        \global\advance #3 by 1
        \expandafter \xdef \csname#1\endcsname {\number #3}%
        \let \newlabelfcn=#4%
        \ifx \newlabelfcn\relax \else
            \expandafter \newlabelfcn \csname#1\endcsname {#1}%
        \fi
    \fi
    \xdef \label  {\csname#1\endcsname}%
    \gdef \suffix {#2}%
    \ifx \suffix\empty \else
        \xdef \suffix {\expandafter\trimspace \suffix\end}%
        \xdef \label  {\label\suffix}%
    \fi
    }%
%
%
%
\newcount\appendixnumber        \appendixnumber=0
\newtoks\appendixstyle          \appendixstyle={\Alphabetic}
\newif\ifappendixlabel          \appendixlabelfalse
\ifundefined{numberedchapters}  \fi
\def\APPEND#1{\par\penalty-300\vskip\chapterskip\spacecheck\chapterminspace
        \global\chapternumber=\number\appendixnumber
        \global\advance\appendixnumber by 1
        \chapterstyle\expandafter=\expandafter{\the\appendixstyle}
\chapterreset
        \titlestyle{Appendix\ifappendixlabel~\chapterlabel\fi.~ {#1}}
        \nobreak\vskip\headskip\penalty 30000}
%

%
%
%
\def\references#1{\par\penalty-300\vskip\chapterskip\spacecheck
        \chapterminspace\line{\fourteenrm\hfil References\hfil}
        \nobreak\vskip\headskip\penalty 30000\reflist{#1}}
\def\figures#1{\par\penalty-300\vskip\chapterskip\spacecheck
        \chapterminspace\line{\fourteenrm\hfil Figure Captions\hfil}
        \nobreak\vskip\headskip\penalty 30000\figlist{#1}}

%
%
\newif\ifdraft\draftfalse
\newcount\yearltd\yearltd=\year\advance\yearltd by -1900
\def\draft{\ifdraft\relax\else\drafttrue
        \def\draftdate{preliminary draft:
                \number\month/\number\day/\number\yearltd\ \ \hourmin}%
        \paperheadline={\hfil\draftdate} \headline=\paperheadline
        {\count255=\time\divide\count255 by 60 \xdef\hourmin{\number\count255}
                \multiply\count255 by-60\advance\count255 by\time
                \xdef\hourmin{\hourmin:\ifnum\count255<10 0\fi\the\count255} }
        \message{draft mode}\fi }
%

%

\def\lnfill{$\mathord- \mkern-6mu\cleaders\hbox{$\mkern-2mu
     \mathord- \mkern-2mu$} \hfill \mkern-6mu \mathord-$}
\def\addline#1{\vbox{\ialign{##\crcr\lnfill\crcr
     \noalign{\kern-9pt\nointerlineskip}$\hfill\displaystyle{#1}\hfil$\crcr}}}

\def\dal{{{\vcenter{\hrule height1pt
     \hbox{\vrule width1pt height7pt \kern7pt \vrule width1pt}
	\hrule height1pt}}\,}}

\def\curlD{{\cal D}}

\def\curlS{{\cal S}}

\def\curlU{{\cal U}}

\def\tr{{\rm tr}\,}
\def\str{{\rm str}\,}
\def\bbZ{\hbox{Z{\kern-2.7pt}Z}}
\def\inZ{{\scriptstyle\in\rm Z{\kern-2pt}Z}}
\def\dal{{{\vcenter{\hrule height1pt
	\hbox{\vrule width1pt height7pt \kern 7pt \vrule width 1pt}
	\hrule height 1pt }}\,}}
\def\ppint{\;\hbox{\bf --}\hbox{\kern-10pt $\displaystyle\int$}\;}
%
%
\def\Aslash{\not{\hbox{\kern-3pt $A$}}}
\def\Bslash{\not{\hbox{\kern-3pt $B$}}}
\def\Cslash{\not{\hbox{\kern-3pt $C$}}}
\def\Dslash{\not{\hbox{\kern-3pt $D$}}}
\def\Eslash{\not{\hbox{\kern-3pt $E$}}}
\def\Fslash{\not{\hbox{\kern-3pt $F$}}}
\def\Gslash{\not{\hbox{\kern-3pt $G$}}}
\def\Hslash{\not{\hbox{\kern-3pt $H$}}}
\def\Islash{\not{\hbox{\kern-3pt $I$}}}
\def\Jslash{\not{\hbox{\kern-3pt $J$}}}
\def\Kslash{\not{\hbox{\kern-3pt $K$}}}
\def\Lslash{\not{\hbox{\kern-3pt $L$}}}
\def\Mslash{\not{\hbox{\kern-3pt $M$}}}
\def\Nslash{\not{\hbox{\kern-3pt $N$}}}
\def\Oslash{\not{\hbox{\kern-3pt $O$}}}
\def\Pslash{\not{\hbox{\kern-3pt $P$}}}
\def\Qslash{\not{\hbox{\kern-3pt $Q$}}}
\def\Rslash{\not{\hbox{\kern-3pt $R$}}}
\def\Sslash{\not{\hbox{\kern-3pt $S$}}}
\def\Tslash{\not{\hbox{\kern-3pt $T$}}}
\def\Uslash{\not{\hbox{\kern-3pt $U$}}}
\def\Vslash{\not{\hbox{\kern-3pt $V$}}}
\def\Wslash{\not{\hbox{\kern-3pt $W$}}}
\def\Xslash{\not{\hbox{\kern-3pt $X$}}}
\def\Yslash{\not{\hbox{\kern-3pt $Y$}}}
\def\Zslash{\not{\hbox{\kern-3pt $Z$}}}
\def\kslash{\not{\hbox{\kern-3pt $k$}}}
\def\dslash{\not{\hbox{\kern-2pt $\partial$}}}
\def\pslash{\not{\hbox{\kern-2.3pt $p$}}}
\catcode`\@=11 
\def\iftpub{\afterassignment\iftp@b\toks@}
\def\iftp@b{\edef\n@xt{\Pubnum={UFIFT--\the\toks@}}\n@xt}
\let\pubnum=\iftpub
\catcode`\@=12 
%
%
%
\overfullrule=0pt
\def\[#1]{[\refrange#1-\end]} 
\def\ghat{{\widehat g}}
\refdef{wigner}{E.~P.~Wigner, {\sl Ann.~Math.}~{\bf 62} (1955) 548;
	{\bf 65} (1957) 203}
\refdef{dyson}{F.J.~Dyson, {\sl J.~Math.~Phys.}~{\bf 3} (1962) 140, 157, 166}
\refdef{penner}{R.C.~Penner, {\sl J.~Diff.~Geom.}~{\bf 27} (1988) 35}
\refdef{BIPZ}{E. Br\'ezin, C. Itzykson, G. Parisi and J.B.~Zuber,
	{\sl Commun.~Math.\hfill\break Phys.}~{\bf 59} (1978) 35}
\refdef{BIZ}{D. Bessis, C.~Itzykson and J.B.~Zuber, {\sl
Adv.~Appl.~Math.}~{\bf%
1} (1980) 109}
\refdef{BK}{E.~Brezin and V.A.~Kazakov, {\sl Phys.~Lett.}~{\bf B236} (1990)
144}
\refdef{GM}{D.J.~Gross and A.A.~Migdal, {\sl Phys.~Rev.~Lett.}~{\bf 64}
	(1990) 127;\hfill\break
	{\sl Nucl.~Phys.}~{\bf B340} (1990) 333}
\refdef{DS}{M.R.~Douglas and S.H.~Shenker, {\sl Nucl.~Phys.}~{\bf B335}
	(1990) 635}
\refdef{kdv}{M.~Douglas, {\sl Phys.~Lett.}~{\bf 238B} (1990); \hfill\break
	      E.J.~Martinec, U.~Chicago preprint EFI--90--62;\hfill\break
		A.~Gerasimov, A.~Markashov, A.~Mironov, A.~Morozov and
		A.~Orlov, \hfill\break Moscow preprint Print--90--0576}
\refdef{skdv}{P.~DiFrancesco, J.~Distler and D.~Kutasov, {\sl Mod.~Phys.~Lett.}
{\bf A5}\hfill\break (1990) 2135;\hfill\break
M.~Awada, U.~Florida preprints UFIFT--HEP--90--18 and --29}
\refdef{mehta}{M.L.~Mehta, {\sl Random Matrices} (Academic Press, 1967)}
\refdef{dewitt}{B.~DeWitt, {\sl Supermanifolds} (Cambridge
	University Press, 1984)}
\refdef{PS}{V.~Periwal and D.~Shevitz, {\sl Phys.~Rev.~Lett.}~{\bf 64} (1990)
	1326; \hfill\break {\sl Nucl.~Phys.}~{\bf B344} (1990) 731}
\refdef{DV}{J.~Distler and C.~Vafa, {\sl Mod.~Phys.~Lett.}~{\bf A5} (1990)
2135}
\refdef{nukes}{J.L.~Rosen, J.S.~Desjardins, J.~Rainwater and W.W.~Havens, Jr.,
 {\sl Phys.~Rev.}\break{\bf 118} (1960) 687}
\refdef{plasma}{C.~Klingshirn and H.~Haug, {\sl Phys.~Rep.}~{\bf 70} (1981)
315}
\refdef{vaz}{C.~Vaz, U. Cincinnati preprint UCTP--102--91}
\refdef{agm}{L.~Alvarez-Gaum\'e and J.L.~Man\~es, CERN preprint
	CERN--TH. 6067/91}
\refdef{gp}{G.~Gilbert and M.J.~Perry, U.~Maryland preprint UMDEPP 91-211}
\pubnum={HEP--91--12}
\date{May 1991}
\titlepage
\title{{\seventeenrm Supermatrix Models}}
\author{Scott A. Yost\foot{Work supported in part by the
Department of Energy, contract DE-FG05-86ER-40272. Address after September 1,
1991: Department of Physics and Astronomy, University of Tennessee,
Knoxville, TN 37996. }}
\address{Department of Physics\break
University of Florida, Gainesville, FL 32611 }

\abstract{ Random matrix models based on an integral over supermatrices
are proposed as a natural extension of bosonic matrix models. The subtle
nature of superspace integration allows these models to have very different
properties from the analogous bosonic models. Two choices of integration
slice are investigated. One leads to a perturbative structure which is
reminiscent of, and perhaps identical to, the usual Hermitian matrix models.
Another leads to an eigenvalue reduction which can be described by a two
component plasma in one dimension. A stationary point of the model is
described. }

\endpage

\chapter{Introduction}

Integrals over random matrices have
had a variety of physical and mathematical applications.
They were originally proposed as a statistical model for studying
the distribution of energy levels of highly excited states of
nuclei\[wigner,dyson,mehta].
The matrices are taken to be from some ensemble of diagonalizable matrices,
usually Hermitian or unitary.
The perturbative expansion of the free energy was found to generate
Feynman rules which are
useful in solving numerous
graph-counting problems\[BIPZ,BIZ,penner]. More recently, these
properties have been found to make certain scaling limits of matrix models
useful for studying two-dimensional gravity, or string theory in ``less than
one dimension\[BK,DS,GM,PS,DV].''

It is natural to consider the extension of these models to supermatrices,
which represent the linear transformations of a vector space having both
even (bosonic) and odd (fermionic) coordinates. All of the algebraic and
analytic methods needed to define ordinary matrix models have extensions
to super\-matrices\rlap.\foot{A general reference, and the source of all
terminology and conventions used here, is ref.~\[dewitt].}
Moreover, supermatrices come in two types: $c$-type, which preserve the
coordinate type of a vector, and $a$-type, which interchange fermions and
bosons. This suggests that a model which includes both could be useful in
defining subcritical superstrings.

This paper will concentrate on $c$-type matrix models, since these are the
most closely analogous to the ordinary models. In particular, ``almost all''
$c$-type matrices are diagonalizable ($a$-type ones need not be),
and diagonalization is a powerful tool in rendering bosonic matrix models
tractable. Unfortunately, the superunitary angular
integral generally does not decouple, so some nonperturbative
methods, such as the orthogonal polynomial method, will lose their power.

Since a $c$-type supermatrix can be decomposed into
a two by two block matrix, with bosonic matrices on the diagonal and fermionic
ones off it, one may suspect that they already contain enough ingredients to
define a discretized superstring model. However, this seems unlikely to
be a proper interpretation, since the potential contains only bosonic
coupling constants,
and since the eigenvalues are all bosonic. If one thinks of $c<1$ strings
as integrable systems\[kdv], what is wanted is a superintegrable
system\[skdv], which
requires a model with fermionic coupling constants to generate both even and
odd $KdV$-type flows.

One of the greatest distinctions between supermatrix models
and ordinary ones is that superspace integration is inherently
ambiguous. There is no geometrically natural reason to restrict the bosonic
components of the supermatrix to be pure real or complex numbers. In general,
they are simply even elements ($c$-numbers) in a Grassmann algebra,
consisting of a {\it body} which is an ordinary number,
and a {\it soul} made of even products of anticommuting numbers\[dewitt].
To define the integration domain
requires specifying a slice through soul-space, and different choices really
should be thought of as different models, with different properties.

Section 2 of this paper describes the general properties of supermatrix
models. Section 3 describes the case when the even entries in the matrix are
ordinary numbers. This choice gives the simplest perturbative expansion,
which turns out to be very similar (perhaps identical) to that of an ordinary
bosonic model. Section 4 describes integrals over matrices which are
constrained
to have pure real eigenvalues with no soul. (These are called
{\it physical} supermatrices\[dewitt].)
When gauge-fixed,  such a model can be described
by a physical system analogous to the Dyson gas of bosonic models.
In this case, the physical analog is found to be a two-component plasma
in one dimension. Although the technology for solving such a matrix model is
presently limited, a saddle point evaluation based on interacting
dipoles is used to illustrate some features of the model.

Three very recent supermatrix model references may be of interest. A paper
by C.~Vaz\[vaz] analyzes a special case of the antisymmetric supermatrix
model. One by L.~Alvarez-Gaum\'e and J.~Man\~es, received after the
completion of the present work, compares the Hermitian
matrix and supermatrix models. Finally, G.~Gilbert and M.J.~Perry\[gp]
propose a quenched $c$-type supermatrix model which appears to
incorporate genuine supersymmetry.

\chapter{Supermatrix Integration}

Supermatrix models can be defined by taking any bosonic model and replacing
all of the ingredients by their superspace analogs. In particular, consider
a Hermitian $c$-type supermatrix\[dewitt] (the only type to be
used in this paper)
$$
	M_{ij} = \pmatrix{ A_{\mu\nu}            & B_{\mu\beta}\cr
			      B^{\dag}_{\alpha \nu} & C_{\alpha\beta} }
\(blockform)
$$
where $A$ and $C$ are bosonic and Hermitian, and $B$ is fermionic.
The notation indicates that $i, j, \ldots$ denote a general index,
while $\mu,\nu,\ldots$ are bosonic and $\alpha, \beta,\ldots$ are
fermionic. If $(-1)^i$ is $+1$ for bosonic indices and $-1$ for fermionic
ones, then the supertrace is defined to be
$$
\str M = \sum_i (-)^i M_{ii} = \tr A - \tr C\;.
\(supertrace)
$$
For an arbitrary potential $V(M)$, the random Hermitian supermatrix model
is defined by the partition function
$$
Z_{mn}(\beta) = \int_{\curlS} dM\ \exp\bigl(-\beta\; \str V(M)\bigr)\;,
\(defint)
$$
where $M$ is an $(m|n)$ supermatrix
acting on vectors with $m$ commuting and $n$ anticommuting components,
and
$$
dM = \prod\; dA_{\mu\nu}\; dC_{\alpha\beta}\; d^2 B_{\mu\beta}\;.
\(measure)
$$
The measures for $A$ and $C$ are the usual linear Hermitian measures.
The integration domain $\curlS$ is chosen so that the bodies of $A$ and $C$
range over the ordinary Hermitian matices, but the souls, which can contain
even products of $B$ components, will be specified later.
Since the supertrace is indefinite, the integral
\(defint) will not actually exist for polynomial potentials. However, this
is a technical problem that can be avoided by considering superunitary
matrices instead, or by multiplying the potential by $i$ and inserting
convergence factors.

When $\curlS$ is chosen such that the Gaussian integral (quadratic potential)
can be evaluated, more general potentials can be handled by a perturbative
expansion. This is true in particular when $A$ and $C$ are taken to be
ordinary soul-less Hermitian matrices, the case considered in section
3. The present section will consider methods which may be applied to more
general cases.

In the bosonic models, the most powerful methods rely on diagonalization
to obtain an integral over eigenvalues alone, factoring out the degeneracy
due to the unitary angular integration\[mehta, BIZ].
Therefore, it is of interest to see
what may be gained by diagonalizing the supermatrix.
A $c$-type hermitian supermatrix can be diagonalized by a superunitary
transformation, except in certain singular cases. The singular cases occur
when one of the eigenvalues of $A$ has the same body as an eigenvalue of $C$.
In that limit, a pair of eigenvectors become bodyless, and cannot be part of
an orthonormal basis, so the diagonalization cannot be carried out. However,
this occurs only at isolated points in the integral.

Assume now that
$$
\curlS = \curlD^{\;\curlU} = \bigl\{ U^{\dag} \curlD U \big| U\in\curlU\bigr\}
\(diagform)
$$
where $\curlU$ is $U(m|n)$, with a possibly unusual choice of souls.
The body of $\curlD$ is just the set of $(m|n)$ real diagonal matrices, and
all off-diagonal elements vanish.
Following the classic matrix-model method\[mehta,BIZ], the
integrand of \(defint) may be multiplied and divided by
$$
\Delta^{-1}(M) =\int_{\curlU} dU\;\prod_{i>j}\delta^2\bigl(
UMU^{\dag}\bigr)_{ij}\;.
\(weight)
$$
Changing variables $ M \rightarrow M^U = U^{\dag}MU$
and using the invariance of the measure and supertrace leads to
$$
Z_{mn}(\beta) =  \int_{\curlU} dU \int_{\curlS^U} d\Lambda\; \Delta(\Lambda)
\exp\bigl( -\beta\;\str V(\Lambda)\bigr)
\(reduction)
$$
where $\Lambda$ is the diagonal matrix of eigenvalues $\lambda_i$ of $M$.

An important difference from the bosonic case is immediately clear.
The integral over the supergroup does not factorize. The choice of souls
for the eigenvalues depends on the group element $U$. If this were a
one-dimensional integral, the choice of souls would not have mattered,
up to a surface term\rlap.\foot{See p.~7 of \[dewitt] for a proof.}
But for multiple
integrals, the dependence is nontrivial. Also, the evaluation of \(weight)
depends on $\curlU$, which is algebraically $U(m|n)$, but may have unusual
soul geometry.

A case where \(reduction) can be evaluated may be obtained by restricting
the eigenvalues to be pure real numbers, and $\curlU$ to be the standard
$U(m|n)$ with its Haar measure. Then the set $\curlS$ consists of what
deWitt\[dewitt] calls the ``physical supermatrices\rlap.''
This case will be the subject of section 4. It has the drawback that
the constraint of having real eigenvalues cannot easily be included
in \(defint), so that even the Gaussian integral becomes complicated,
and perturbative Feynman rules are difficult to develop.

A possible compromise would be to diagonalize only $A$ and $C$, by performing
a superunitary transformation $U\times V \in U(m)\times U(n)$. Then
$B\rightarrow UBV^{\dag}$. The Jacobians for $A$ and $C$ are the usual
Vandermonde determinants, while the Jacobian for $B$ is just 1, so
$$
Z_{mn}(\beta) = \int \prod da_\mu dc_\alpha dB_{\nu\beta}
\prod_{\mu<\nu}(a_\mu-a_\nu)^2\prod_{\alpha<\beta}(c_\alpha-c_\beta)^2
\exp\bigl(-\beta\;\str V(M)\bigr)
\(partdiag)$$
where $M$ is now defined with $A = {\rm diag}(a)$, $C = {\rm diag}(c)$, and
the volume of $U(m)\times U(n)$ has been divided out.
The remaining integral over $B$ can be replaced by an integral over a
$U(m|n)/U(m)\times U(n)$ coset, if desired.  It is not clear whether
\(partdiag) is a useful supplement to \(defint). This question will not be
persued here.

A simple example may be helpful to clarify the issues in this section.
Consider the case $m=n=1$. Then $A$ and $C$ become real even supernumbers
$a$ and $c$, while $B$ becomes a complex odd Grassmann number $\beta$.
The eigenvalues of $M$ are
$$
\lambda = a + (a-c)^{-1} \beta\beta^{*},\qquad
\mu     = c + (a-c)^{-1} \beta\beta^{*},
\(eigenvalues)
$$
and $UMU^{\dag}$ is diagonal when the superunitary matrix is
$$
	U = \pmatrix{ 1-\coeff12 \alpha\alpha^{*} & \alpha\cr
			-\alpha^{*}	& 1+\coeff12 \alpha\alpha^{*}\cr }
\(superU)
$$
with $\alpha = (a-c)^{-1}\beta$.

The partition function will now have the form
$$\eqalign{
Z =& \int da\,dc\,d\beta d\beta^{*} \exp\bigl(-\str V(M)\bigr)\cr
  =& \int d\alpha d\alpha^{*} \int d\lambda d\mu (\lambda-\mu)^{-2}
		\exp\bigl(-V(\lambda) + V(\mu)\bigr).
}\(11part)
$$
This is a special case of \(reduction) with $dU = d\alpha d\alpha^{*}$
and $\Delta(\lambda,\mu) = (\lambda-\mu)^{-2}$. The unimportant diagonal
subgroup of $U(1|1)$ is omitted. Since no even parameters are
needed in $U$, there is no ambiguity in expressing the group manifold
$\curlU$ in this case. The supernumbers $a$ and $c$
can be decomposed into body and soul, which have the form
$$\eqalign{
a =& a_B + a_S = a_B + f(a_B, c_B) \beta\beta^{*} \cr
c =& c_B + c_S = c_B + g(a_B, c_B) \beta\beta^{*}, \cr
}\(bodysoul)
$$
where $a_B$ and $c_B$ range over the real numbers, and $f$ and $g$ are ordinary
real-valued functions. This form is unique if the souls depend only on $\beta$
and $\beta^{*}$.

The integral can be reduced to an integral over the bodies
by including a Jacobian factor
$$
\int da\,dc = \int_{-\infty}^{\infty} da_B dc_B \bigl[ 1 + (f_{,1} + g_{,2})
	\beta\beta^{*}\bigr].
\(souljac)
$$
The presence of such factors can make a direct evaluation of \(defint)
complicated for large matrices, even for the simplest potentials. It is
useful to choose $\curlS$ so that either \(defint) or \(reduction) is as
simple as possible. These two cases are the subject of the following
sections.

\chapter{Ordinary Hermitian Supermatrix Model}

The partition function \(defint) can be expressed most simply when $A$ and $C$
are chosen to be ordinary soul-less Hermitian matrices. This case will be
referred to as the ``ordinary Hermitian supermatrix model\rlap.'' No
troublesome Jacobian factors of the form \(souljac) complicate the evaluation,
so the Gaussian integral can be evaluated exactly. More complicated
potentials can be handled perturbatively, by developing Feynman rules\[BIZ].
The identification of dense Feynman diagrams with Riemann surfaces is the
reason for the relevance of a scaling limit of ordinary matrix models
to two-dimensional gravity\[BK,DS,GM]. Therefore, it should be
expected that the
connection to gravity can be made most directly for the ordinary Hermitian
supermatrix model.

Consider a quadratic potential $\coeff12 \str M^2$. The Feynman
rules and Wick's theorem can be derived as in the usual case, by calculating
the expectation value of a source $\exp(\str(JM))$, and varying with
respect to $J$ to obtain the expectation values needed for a perturbative
expansion. The propagator is simply
$$
\vev{M_{ij} M^{*}_{kl}} = (-)^j \delta_{ik}\delta_{jl}.
\(propagator)
$$
Wick contractions must always be carried out after permuting the matrices
so that they are adjacent, keeping track of the signs $(-)^{(i+j)(k+l)}$
introduced by commuting $M_{ij}$ past $M_{kl}$. Symmetry factors can be
calculated as in the usual matrix models, following the rules of ref.~\[BIZ].

\insertfigure{two-loop}{\vskip 1.5 truein}{The three lowest-order graphs
for the quartic model.}
$$
V(M) = \coeff12 \str M^2 + g\; \str M^4\;,
\(quartic)
$$
the lowest-order contribution to the free energy comes from the
two-loop graphs in figure 1, which represent $\vev{\str M^{4}}$.
Evaluating them gives
$$\eqalign{
g\vev{\str M^4} =& g\sum_{i,j,k,l} (-)^i \bigl[ \vev{M_{ij}M^{*}_{kj}}
		\vev{M_{kl}M^{*}_{il}} + (-)^{(j+k)(k+l)}\vev{M_{ij}
		M^{*}_{lk}} \vev{M_{jk}M^{*}_{il}} \cr
&\qquad \qquad + (-)^{(j+l)(l+i)} \vev{M_{ij}M^{*}_{il}}\vev{ M_{jk}M^{*}_{lk}}
			\bigr]\cr
=& g \sum_{i,j,k,l} \bigl[ (-)^{i+j+l} \delta_{ik}\delta_{jj}\delta_{ki}
	\delta_{ll} + (-)^{i+j+k}(-)^{(j+k)(k+l)}\delta_{il}\delta_{jk}
	\delta_{ji}\delta_{kl} \cr
&\qquad \qquad + (-)^{i+j+k}(-)^{(j+l)(l+i)}\delta_{ii}
	\delta_{jl}\delta_{jl}\delta_{kk}\bigr]\cr
=& g \bigl[ (m-n)^3 + (m-n) + (m-n)^3\bigr].\cr
}\(twoloop)
$$
The three contributions to \(twoloop)
come respectively from the three graphs in figure 1. Graphs (a) and (c)
are planar, while (b) is nonplanar. It was possible to rescale $g$ so that
\(twoloop) is identical to the standard Hermitian matrix result\[BIZ].
If $g = {\widehat g}(m-n)^{-1}$, then \(twoloop) may be written
$$
{\widehat g}(m-n)^{-1} \vev{\str M^4} = {\widehat g} \bigl[2(m-n)^2+1\bigr]
\(rescaled)
$$
The power of $(m-n)^2$ is $(1-G)$, where $G$ is the genus of the surface
identified with the graph by filling in faces in the usual manner.

\insertfigure{three-loop}{\vskip 2 truein}{The inequivalent connected
three-loop graphs with two quartic vertices.}
This provides evidence that the supermatrix model will have a $2d$ gravity
interpretation, when $m,n \rightarrow\infty$ with $n/m$ fixed. A computation
of all three-loop connected graphs containing two quartic vertices provides
further
evidence for this. The total contribution of these graphs, shown in figure 2,
to the free energy is
$$
{{\widehat g}\over 2!(m-n)^2}\vev{(\str M^4)^2}_{\rm conn.} =
	{\widehat g} \bigl[18 (m-n)^2 + 30\bigr]\;,
\(threeloop)
$$
which is identical to the ordinary Hermitian matrix
result\[BIZ], with $N$ replaced by $m-n$. In fact, it is clear that in
an arbitrarily complex diagram, each closed index loop (face in a generalized
triangulation of the surface) will give a
factor of $m\pm n$. If the pattern suggested by low order computations
persists, then only $m-n$ will occur, and the model will be {\it precisely}
equivalent to the $|m-n|\times|m-n|$ Hermitian matrix model.
(Ref.~\[agm] provides some further evidence that this is the case.) However, if
any factors of $m+n$ occur, the topological expansion will still go through,
but the weighting factors for various surfaces will be changed. More
powerful combinatorial methods
are needed to settle this question\rlap.

If the identification with the $|m-n|$ Hermitian model persists, then this
may suggest a connection with the bosonic sector of a subcritical
superstring theory. Ordinary superstrings have a bosonic Neveu-Schwarz sector
which by itself is closely analogous to the purely bosonic string, in
the sense that is can contain the same massless spacetime fields (gravity,
\etc)
with the same low-energy effective action. The fermions play a purely internal
role in this sector. This seems analogous to the role played by the fermions
in a $c$-type supermatrix. They play an internal role, but in a ``spacetime''
sense, the model behaves as a bosonic one. Adjoining $a$-type supermatrices
could provide an analog of a Ramond sector, containing true fermions.
Quenching, or constraining some elements of a supermatrix element, may
also lead to a supersymmetric theory\[gp].

\chapter{Physical Supermatrix Model}

Having considered the case where the even entries in the supermatrix are
pure numbers, it is now natural to turn to the other relatively
simple case, where the eigenvalues of the supermatrix are pure real numbers.
Such Hermitian supermatrices may be thought of as observables in
a quantum-mechanical system defined on a supermanifold\rlap.\foot{Such systems
are described in chapter 5.3 of ref.~\[dewitt].}

Incorporating the constraint that $M$ have real eigenvalues directly into
\(defint) is complicated. A Jacobian factor will be needed, which is
difficult to compute in general. However, working with the eigenvalue
reduction \(reduction) is simple in this case. The eigenvalues $\lambda_i$
are integrated over the ordinary real numbers, and the supergroup integral
over $U(m|n)$ factorizes. In fact, that integral vanishes, as can be
seen from \(11part) for the $(1|1)$ supermatrix example, or in general
from the fact that \(defint) vanishes when $\partial\lambda_i/\partial
B_{\mu\alpha}=0$. In any case,
the supergroup integral will be dealt with by fixing the gauge,
and simply dropping it. This gives
the closest analog of the ordinary matrix model eigenvalue reduction.
The remaining eigenvalue partition function will be essentially a vector
model whose form and symmetries are inherited from the underlying
supermatrix model.

The only ingredient still needed in \(reduction) is the determinant $\Delta$.
This can be calculated by representing $U$ in \(weight) as $e^A$ with $A$
anti-Hermitian. Then, if $\Lambda$ is the diagonalized matrix,
$$
\bigl( U\Lambda U^{\dag} \bigr)_{ij} = A_{ij}(\lambda_j-\lambda_i) + \ldots
\(Uexp)
$$
and, keeping track of the Grass\-mann character of the delta functions in
\(weight), the determinant is found to be
$$
\Delta(\lambda) = \prod_{\mu<\nu}(\lambda_\mu - \lambda_\nu)^2
		  \prod_{\alpha<\beta}(\lambda_\alpha - \lambda_\beta)^2
		  \prod_{\mu,\alpha} (\lambda_\mu - \lambda_\alpha)^{-2}
\(det)
$$
in the index conventions of \(blockform).
The poles occur at supermatrices which technically are not diagonalizable
for the reasons noted in section 2.
The gauge-fixed physical supematrix model is then defined by the equation
$$
Z_{mn}(\beta) = \int \prod_i d\lambda_i \Delta(\lambda) \exp -\beta\Bigl(
\sum_{\mu=1}^m V(\lambda_\mu) - \sum_{\alpha=1}^n V(\lambda_\alpha)\Bigr).
\(physical)
$$

This model is clearly inequivalent to the usual hermitian model, due to
the presence of a denominator in \(det). The ordinary matrix models were
introduced as a model for describing the eigenvalue distribution of a
random physical operator. In that case, Wigner\[wigner] observed that
the eigenvalues repel
each other, and Dyson\[dyson] noted that they can be interpreted as a
gas of charged particles with $e^2 = 1/\beta$
in one dimension, at temperature $1/\beta$. In the limit $N,\beta\rightarrow
\infty$ with $\beta<N$, the gas freezes into a crystal with a charge
distribution
determined by the external potential $V$. The critical case takes
$\beta/N\rightarrow 1$, so that fluctuations about the crystal become
important, and in the perturbative expansion, higher genus graphs contribute
comparably to planar ones\[BK,DS,GM].

The physical analog for the supermatrix model is a two-component plasma
in one dimension, consisting of equal and opposite charges $\pm\beta^{-1/2}$
at temperature $1/\beta$, in an external potential $V$. The $\lambda_\mu$
may be considered to be $m$ positive charges, while the $\lambda_\alpha$
are $n$ negative charges. At very low temperatures, the positive
charges will pair with a negative partner, leading to an effective
theory of $n$ dipoles interacting with $m-n$ positive charges when $m>n$.

Assume $m>n$ and introduce variables
$$
x_i = \coeff12(\lambda_i + \lambda_{m+i}),\qquad u_i =
\coeff12(\lambda_i - \lambda_{m+i}),\qquad y_a = \lambda_{n+a}
\(newvars)
$$
for $i = 1,\ldots,n$, and $a = 1,\ldots, m-n$.
In the limit where the first $n$ positive and negative charges pair,
$x_i$ is a center of mass coordinate for the $i$th dipole, $u_i$ is its
charge displacement, and $y_a$ labels the $m-n$ left-over positive
charges. In terms of these variables, \(physical) may be rewritten
$$\eqalign{
Z_{mn} = 2^n \int &\prod_{i=1}^{n} {dx_i du_i\over u_i^2} \prod_{a=1}^{m-n}
dy_a \prod_{i\ne j} \left[ {1 - (u_i-u_j)^2(x_i-x_j)^{-2} \over
			    1 - (u_i+u_j)^2(x_i-x_j)^{-2}}\right]^2\cr
	\times & \prod_{i,a} \left[ {1 + u_i(x_i-y_a)^{-1} \over
		1 - u_i(x_i-y_a)^{-1}} \right]^2
	\prod_{a,b} \bigl(y_a-y_b\bigr)^2\cr
\times & \exp-\beta\left\{ \sum_a V(y_a) + \sum_i \bigl[
	V(x_i+u_i)-V(x_i-u_i)\bigr]\right\}.\cr
}\(newpartition)
$$

Clearly, $Z_{mn}$ is dominated by configurations where $u_i\rightarrow 0$,
and by configurations equivalent to this up to index permutations. These
configurations actually cause $Z$ to diverge, which is a price for simply
dropping the $U(m|n)$ integral. However, once the internal energy of the
dipoles is regulated by cutting off the displacements, no further divergences
occur, provided the potential is well-chosen.

For simplicity, further attention will be restricted to the neutral plasma,
with $m=n$. In the dipole approximation,
cutting off the displacements at $\delta_i$ gives
$$\eqalign{
Z \cong &\  2^n n!\int_{-\infty}^{\infty} d^n x \int_{\delta_i} {du_i\over
u_i^2}\cr
&\ \times\sum_{\epsilon_i =\pm1}
\exp-\beta\left\{ \sum_i \epsilon_i\bigl[V(x_i+u_i)-V(x_i-u_i)\bigr]
	+ \sum_{i\ne j} \epsilon_i\epsilon_j V_{ij} \right\}\cr
}\(cutZ)
$$
where $u_i$ are now considered to be small and positive
and the dipole interaction is
$$
V_{ij} =
       =\ -{4\over\beta} \tanh^{-1} \left[ {2u_i u_j\over (x_i-x_j)^2
			- u_i^2-u_j^2} \right]
\ \cong\ -{8\over\beta} {u_i u_j\over (x_i-x_j)^2}\;.
\(interaction)
$$
To lowest order in $u$, the interaction is attractive, but the exact form
shows that higher order repulsive effects will prevent a singularity
in $V_{ij}$ for $x_i\rightarrow x_j$. The dipole interaction with the external
potential is $\epsilon_i u_i V'(x_i)$, where $V'$ is the derivative of
$V$ with respect to $x$. Therefore, positively polarized dipoles are
drawn to the minima of $V'$ while negative ones are drawn to its maxima.

It is informative to choose a simple potential, and find the behavior
of the free energy near a saddle point.
Consider the example of a cubic potential
$$
\beta V(x) = \coeff12 x^2 + \coeff13 \beta^{-1/2}g x^3.
\(cubic)
$$
In this case, the true minima of the potential should come from
unstable configurations where negative dipoles lump together and run off
to infinity, or positive ones all condense at $x = -\beta^{1/2}g^{-1}$.
The former case will be neglected, since it can be tamed by modifying the
potential at large $|x|$ (a situation familiar in matrix models).
The latter case is important, but requires the inclusion of the higher-order
repulsive effects in $V_{ij}$ for meaningful results.

Therefore, an
unstable stationary point, where negative dipoles are repelled from the
minimum of $V'$ but attracted to each other, will be considered for
illustrative purposes. (This is opposite the usual situation for matrix
models.)
It is a simple enough case to allow a direct
comparison with the saddle point analysis of the standard matrix
model\[BIPZ], and it is useful to know exactly how the supermatrix analysis
differs.

In the $n,\beta\rightarrow\infty$ limit, it is useful to
define a continuum variable $x(t)=n^{-1/2}x_i $ with $t= i/n$.
Near the saddle point, the dipole
approximation should be valid, and we will work with the effective
dipole energy
$$
n^2\beta E = 2\beta\int_0^1 dt\;p(t)V'(x(t)) - 8 \int_0^1\int_0^1 dt\,dt'\;
{p(t)p(t')\over \bigl(x(t)-x(t')\bigr)^2} + {2\over n}\int_0^1 dt\;
\log\bigl( n^{1/2} p(t)\bigr).
\(dipenergy)
$$
The continuum dipole is defined to be $p(t) = n^{-1/2}\epsilon_i\delta_i$
in terms of the polarization and cutoff in \(cutZ), and for the cubic
potential,
$$
\beta V'(x(t)) = x(t) + \ghat x^{2}(t), \qquad \ghat = \left({n\over\beta}
	\right)^{1/2}g.
\(cubepot)
$$

Introducing a dipole density $P(x) = p(t) dt/dx$ and dropping the dipole
self-energy term gives
$$
n^2\beta E = 2\int dx\;P(x)\bigl(x+\ghat x^2\bigr) - 8\int\int dx\,dx'\;
	{P(x)P(x')\over(x-x')^2}.
\(ourpot)
$$
The saddle point equation $\delta E/\delta p(t) = 0$ is satisfied when
$\partial_x (\delta E/\delta P(x)) = 0$. For \(ourpot) this implies
$$
x + {1\over 2\ghat} = -{16\over\ghat} \ppint dx' {P(x')\over (x-x')^3}\;,
\(saddle)
$$
using the same principle part prescription as ref.~\[BIPZ] to handle the
coincident point. (Recall that \(cutZ) shows that the coincident limit
is not truly singular.)

Since the saddle point which will be found is the unstable one with
negative dipoles clustered about the minimum of $V'$ (using some foresight),
it is convenient to define the integrated dipole density as
$$
	\int dx P(x) = - p_0.
\(p0)
$$
The quantity $p_0$ is determined by how $n^{-1/2}\delta_i$ is tuned as
$n\rightarrow \infty$, $\delta_i\rightarrow0$, and should be thought of
as a new parameter of the model.
It is also convenient to introduce a new variable
$$
w = \kappa \left( x + {1\over 2\ghat}\right),
\(wdef)
$$
centered at the minimum of $V'$, with scale
$\kappa^{-4} = 32 p_0/\ghat$ chosen for later simplicity.
Then the normalized negative dipole density $\rho$ defined by
$$
P(x) = -\coeff12 \pi\kappa p_0 \rho(w)
\(rhodef)
$$
has total integral $+2/\pi$. (This integral was fixed for later convenience.)

In the new variables, the stationary condition \(saddle) becomes
$$
w = {\pi\over 4}\ppint dw' {\rho(w')\over(w-w')^3}\,,\qquad |w|<a.
\(simplesaddle)
$$
Due to the symmetry of the problem, $\rho(w)$ will be an even function,
with support in an interval $[-a,a]$ to be determined.
The density does not have to be positive, since dipoles can have two
polarizations, but the integral must be $2/\pi$ by construction.
A solution can be found following ref.~\[BIPZ]'s treatment of the
ordinary quartic matrix model, by introducing an analytic function
$$
F(w) = {\pi\over 4}\int_{-a}^a dw'{\rho(w')\over(w-w')^3},
\(anal)
$$
which is analytic for complex $w$ with cut $[-a,a]$, behaves as ${1\over2}
w^{-3}$ for $|w|\rightarrow\infty$, and satisfies
$$
F(w\pm i\epsilon)  = w \mp \coeff{1}{2} \pi i \rho''(w)
\(determiningequation)
$$
for $w$ in $[-a,a]$.

The solution is
$$
F(w) = w - \sqrt{w^{2} - w^{-2}},
\(Fsolution)
$$
which requires $a=1$ and
$$
\rho''(w) = {2\over |w|}\sqrt{1-w^4}
\(2deriv)
$$
for $|w|<1$.
Integrating once gives
$$
\rho'(w) = \left(\sqrt{1-w^4} - \tanh^{-1}\sqrt{1-w^4}\right)\,
\hbox{\rm sgn}\;w .
\(1deriv)
$$
The logarithmic singularity at $w=0$ is integrable, and the resulting
density will be finite. The integration constants are fixed by adding
a term
$$
-h \bigl(1- |w|\bigr)
\(addterm)$$
to $\rho(w)$, with coefficient $h$ chosen such that the integrated density
is $2/\pi$, as required by the defining conditions. Eqn.~\(1deriv)
may be integrated numerically, leading to $h \cong .147$, for which
the final solution $\rho(w)$ is shown in figure 3.

\insertfigure{rhograph}{\vskip 2 truein}{ The normalized negative dipole
density
$\rho(w)$ as a function of the coordinate $w$ centered at the minimum of
the quadratic potential $V'$.}
The energy for this configuration is
found by substituting $\rho$ into \(ourpot). Using some integration by
parts, \(2deriv), \(1deriv), and \(addterm) suffice to obtain $\beta E$.
The double integral can be done with the help of \(saddle) as
in ref.~\[BIPZ], leading to
$$
n^2\beta E = \coeff38 \,\ghat^{\,-1}p_0
\ +\ \coeff13 \pi \,\ghat^{\,1/2}p_0^{3/2}\bigl\{I(\alpha)-\coeff{1}{3}+
h\log(1 - \alpha)\bigr\}
\(Eint)$$
where $\alpha^2 = 2^{9}\,\ghat^{\,3}\, p_0$ and
$$\eqalign{
I(\alpha) = -\int_0^1 {dz\over z}\sqrt{1-z^2}\;\log\bigl( 1-\alpha z\bigr)
	= {\sqrt{\pi}\over2}\sum_{n=1}^{\infty}{\Gamma\left({n\over2}\right)
\over\Gamma\left({n+1\over2}\right)}{\alpha^n\over n(n+1)}\;.
}\(lastint)
$$
The surface interpretation of a matrix model depends only on the non-analytic
behavior of the partition function.
The expression \(Eint) has non-analytic behavior when
$$
	2^9 g^3 p_0 \left({n\over\beta}\right)^{3/2}
			\longrightarrow\ 1\,.
\(nonanal)
$$
This can occur for $n/\beta\rightarrow 1$ if  $g^3 p_0 = 2^{-9}$.

Since this fixed point is unstable, it is not clear that it has any
application in $2d$ gravity.
(However, inverted potentials sometimes come up in scaling
limits of matrix models\[GM], so instabilities are not always as bad
as they seem.) This is the only kind of stationary point that
can be found within the limits of the dipole approximation of the
one-dimensional plasma. However, the interaction potential $V_{ij}$
contains higher-order repulsive effects, so stable fixed points may
eventually be found. A more exact treatment would be desirable,
but this is more difficult than for the usual matrix models, where the
orthogonal polynomial method yields relatively easy exact results.

\chapter{Conclusions}

The motivation for introducing supermatrix models is their apparent
mathematical similarity to ordinary matrix models. This suggests that they
could provide a way to incorporate fermions in a matrix model without
sacrificing integrability.

However, when these models are investigated in detail, it becomes clear that
the nature of superspace gives them very different properties from ordinary
matrix models, and that certain useful tools, especially the orthogonal
polynomial method, are not easily applied. The source of the most unique
features of supermatrix models is the inherent ambiguity in integrating
over even Grassmann variables.

The most obvious definition of the supermatrix integral gives a perturbative
expansion which implies that a scaling limit of the model
will have a $2d$ gravity interpretation (but not supergravity, since the
models considered have only bosonic coupling constants). The fact that
the superunitary integral does not decouple in this case means that the
eigenvalue reduction an orthogonal polynomial method are not readily
applicable, although eventually suitable extensions may be found.

The leading terms in the perturbative expansion of the quartic
$(m|n)$ Hermitian supermatrix model are identical to those of the
$(m-n)$ bosonic Hermitian matrix model.
This may suggest a connection to the ``spacetime'' bosonic sector
of a subcritical superstring, to which fermionic partners must be
adjoined to complete the theory. The fact that supermatrices contain
odd Grassmann variables is not in itself reason to believe that they would
constitute a supersymmetric theory by themself. Perhaps the appropriate
partners are supermatrices with the bosonic and fermionic entries enterchanged.

Since the supermatrix integrand depends only on invariants, it will actually
vanish whenever the superunitary integral does decouple. Nevertheless, it
is useful to define such models by gauge-fixing, since they provide the
closest super-analog of Wigner's model for the eigenvalue distribution
of a physical operator. The result is that the eigenvalues of a physical
super-operator separate into two classes. Within each class, the eigenvalues
repel, as in Wigner's model. However, eigenvalues in each class attract those
of the other. If any real system has such properties, perhaps supermatrices
could provide a useful model for some of its statistical properties, as
matrix models did  with some degree
of success\[nukes] for the highly-excited states of nuclei.

In analogy with the Dyson gas interpretation of matrix models, the
eigenvalues of a supermatrix may be though of in terms of a two-component
plasma in one dimension. This physical analog could provide useful intuition
for solving the matrix model. Conversely, the matrix model, when
sufficiently developed, could provide a new source of information about
such plasmas. Electron-hole plasmas in optically-excited
semiconductors are a well-studied system\[plasma] which may be
somewhat analogous.

The analysis presented here should be considered preliminary.
Assuming that opposite charges condense into pairs at low temperatures,
an effective dipole gas was constructed, and one of the stationary points
for a cubic potential was analyzed. The unstable fixed point
analyzed was chosen because it is the simplest. If a way is found to
deal with the higher-order repuslive effects present in (4.7), it may
be possible to analyze a stable fixed point instead. It may be necessary
to extend the analysis to include
effective neutral particles containing more than two charges, \ie\ higher
multipoles. Developing a more exact treatment of the
plasma could reveal some interesting physics, and perhaps new
$2d$ gravity models.

\ack

I would like to thank M.~Awada,
Z.~Qiu, P.~Ramond, S.-J.~Rey, S.-J.~Sin, and R.P.~Woodard for discussions
and comments.

\references{}
\figures{}
\end